\documentstyle[aps]{revtex}
\begin{document}
\title{Reparametrization invariant statistical inference and 
gravity}
\author{Vipul Periwal}
\address{Department of Physics,
Princeton University,
Princeton, New Jersey 08544}

\def\dd{\hbox{d}}
\def\tr{\hbox{tr}}\def\Tr{\hbox{Tr}}
%\baselineskip=14truept
\maketitle
\begin{abstract} Bialek, Callan and Strong have recently given a
solution of the problem of determining a continuous probability 
distribution from a finite set of experimental measurements by 
formulating it as a one-dimensional quantum field theory.  This 
letter gives a reparametrization-invariant solution of the problem,
obtained by coupling to gravity.  The case of a large number of 
dimensions may involve quantum gravity restricted to metrics of
vanishing Weyl curvature.
\end{abstract}
In many instances in physics, or in other fields, it is necessary to
determine the probability distribution that underlies a finite set
of experimental results involving continuous variables.  From the
practical point of view, one usually has a definite finite-parameter
model for the 
probability distribution on theoretical grounds, and the parameters 
are fixed by fitting to the data set. 
This introduces a theoretical bias, so it is of interest to 
attempt a direct determination of the probability 
distribution without resorting to finite-dimensional approximations.  
Of course, for any finite data set, one obtains only a probabilistic 
description of the probability distribution, but the spread of 
possible probability distributions decreases as the size of the data 
set is increased\cite{history}.

Bialek, Callan and Strong\cite{bialek} 
have recently given an elegant formulation of
this problem in one dimension.  They used Bayes' rule to write the 
probability of the probability distribution $Q,$ given the data 
$\{x_{i}\},$ as
\begin{eqnarray}
P[Q(x)| x_1 , x_2 , ... , x_N]
\,
 = 
{{P[x_1 , x_2 , ... , x_N | Q(x)] P[Q(x) ]}
\over {P(x_1 , x_2 , ... , x_N )}} \, =
{{Q(x_1 ) Q(x_2 ) \cdots Q(x_N) P[Q(x)]}
\over{\int [dQ(x)] Q(x_1 ) Q(x_2 ) \cdots Q(x_N) P[Q( x)]}} ,
\label{conditional}
\end{eqnarray}
where the factors $Q(x_{i})$ arise because each $x_{i}$ is chosen 
independently from the distribution $Q(x),$ and $P[Q]$ encodes the 
{\it a priori} hypotheses about the form of $Q.$     The optimal 
least-square estimate of $Q,$ $Q_{{\rm est}}(x,\{x_{i}\}),$ is then
\begin{equation}
Q_{\rm est} (x; \{x_i\} )
=
{{\langle Q(x) Q(x_1 ) Q(x_2 ) \cdots Q(x_N) \rangle ^{(0)}}\over
{\langle Q(x_1 ) Q(x_2 ) \cdots Q(x_N) \rangle^{(0)}}} ,
\label{est}
\end{equation}
where $\langle \cdots \rangle^{(0)}$ denotes an  expectation value with
respect to the {\it a priori} distribution $P[Q(x)]$.  

In this field-theoretic setting, Ref.~\cite{bialek} assumed that 
the prior distribution $P[Q]$ should penalize large gradients, so
written in terms of an unconstrained variable $\phi \equiv -\ln (\ell 
Q) \in (-\infty,+\infty),$ they assumed 
\begin{eqnarray}
P_\ell[\phi(x)]
 =  {1\over Z} \exp\left[ -{\ell\over2} \int dx (\partial_x \phi)^2
\right]
\, \times
\delta \left[ 1 - {1\over \ell} \int dx e^{-\phi(x)}\right] ,
\end{eqnarray}
with $\ell$ a parameter that they later averaged over independently.  
This form for the prior distribution is very simple, and quite 
minimal in terms of underlying assumptions, so 
conclusions drawn from it should be robust.  There is, however, 
an important aspect in which this prior distribution does not measure 
up, so  at this juncture my analysis deviates 
from Ref.~\cite{bialek}. 

The parameter $\ell$ is a global 
variable, independent of $x$ in the analysis of Ref.~\cite{bialek}, 
whereas there are
some reasons for wishing to keep this scale parameter local, as I now 
discuss.
We expect that the parametrization of the data $\{x_{i}\}$
 should not affect the
probability distribution that we determine from it.  To be more 
precise, we expect that if $f(x)$ is a monotone function of $x,$ then
$\{f(x_{i})\}$ should determine a probability distribution $Q_{f}$ 
which is related to $Q_{x}$ by
\begin{eqnarray}
\int_{x_{-}}^{x_{+}} dx' Q_{x}(x') = \int_{f(x_{-})}^{f(x_{+})} dg 
Q_{f}(g)
\end{eqnarray}
for arbitrary values of $x_{\pm};$
in words, the estimated distributions must be covariant with respect 
to reparametrizations of the data.  The assumed form of $P[Q]$ in
Ref.~\cite{bialek} does not possess this reparametrization-covariance,
so my aim here is to change  their formulation to find a solution that 
does.  The reparametrization invariance of results derived from the 
data is, obviously, high on the list of desiderata for this problem. 

In fact, the resolution of this problem is implicit in 
Ref.~\cite{bialek}, if one takes care to read their discussion 
regarding the determination of $\ell,$ as well as their discussion
regarding the error in determining the correct probability 
distribution from a finite data set.  As observed there, the 
introduction of $\ell$ is really  a determination of a scale on
which variations of $Q$ are viewed as too rapid.  We should not
expect that this length scale is a global length scale, indeed, quite
intuitively one would like to allow derivatives of different 
magnitudes in $Q$ in 
regions where data points cluster or are sparse.  

From a particle theory perspective, these two problems, that of 
reparametrization invariance, and that of a local scale determination,
signal the need to  introduce a metric, in other words, to 
couple the field $\phi$ to gravity.  I shall show in the following 
that this coupling to `quantum' gravity makes the analysis easy, 
leading to a final result that could not take a simpler form.

I write $Q(x) \equiv \sqrt{h(x)} \exp(-\phi(x))/\ell,$
with 
\begin{eqnarray}
P[\phi,h] = {1\over Z} \exp\left[ -{\ell\over2} \int dx 
\sqrt{h(x)}^{{-1}}(\partial_x \phi)^2
\right] \delta \left[ 1 - {1\over \ell} \int dx \sqrt{h(x)} e^{-\phi(x)}\right] ,
\end{eqnarray}
since in one dimension the metric, $h(x),$ is a section of a real line bundle, 
hence in any fixed coordinate system, is just a `function'.  Note that 
$\ell$ can be absorbed in $\sqrt{h(x)},$ so we set $\ell=1$ in the 
following. 
The inverse of the metric is just $1/h(x)$ and the reparametrization  
invariant volume element is $\sqrt{h(x)} dx,$ so $P[\phi,h]$ is 
reparametrization invariant.  Now, we want to evaluate
\begin{eqnarray}
&&\langle Q(x_1 ) Q(x_2 ) \cdots Q(x_N) \rangle^{(0)}
\nonumber\\
&&\,\,\,\,\,
=
\int D\phi {{ Dh}\over{Diff_{+}}}
P[\phi,h] \prod_{i=1}^N \sqrt{h} \exp[-\phi(x_i)] \\
&&\,\,\,\,\, =
 {1\over Z} \int {{d\lambda }\over{2\pi}}
\int D\phi {{ Dh}\over{Diff_{+}}}
\exp\left[ - S(\phi,h; \lambda ) \right] ,
\label{problem}
\end{eqnarray}
where 
\def\ee{{\rm e}}
\begin{eqnarray}
S(\phi,h; \lambda ) =
{1\over2} \int dx \sqrt{h(x)}^{{-1}}(\partial_x \phi)^2 
\,
+i{\lambda } \int dx \sqrt{h(x)} \ee^{-\phi(x)}
+\sum_{i=1}^N [\phi (x_i )- {1\over 2} \ln h(x_{i})] - i\lambda.
\end{eqnarray}
Notice that the integral over all metrics has been divided by 
the volume of the group of orientation preserving diffeomorphisms, as is 
appropriate for a reparametrization invariant action.  In one 
dimension, this division eliminates all but one degree of freedom from
the metric.  Further, there 
is no operational way to distinguish between the factor 
$\sqrt{h(x_{i})}$ and $\exp(-\phi(x_{i}))$---in other words, these 
must occur together in $Q(x_{i}).$ 

The equations of motion that follow from varying $S$ are
\begin{eqnarray} 
&&-{1\over 2} (\phi')^{2} {1\over h} + i\lambda \exp(-\phi) -
\sum_{i} {1\over \sqrt{h(x_{i})}} \delta(x-x_{i}) = 0 \ ,
\label{hvar}
\\
&& - \left({1\over \sqrt{h}}\phi'\right)' 
-i\lambda\sqrt{h}\exp(-\phi) + \sum_{i}\delta(x-x_{i}) = 0 \ ,
\label{pvar}
\\
&& \int dx \sqrt{h} \exp(-\phi) =1\ ,
\label{lvar}
\end{eqnarray}
where I have used primes to denote $d/dx.$
Introduce a variable 
\begin{equation}
y(x) = \int^{x} \sqrt{h(s)} ds,
\end{equation}
then 
it follows from eq.~\ref{hvar} and eq.~\ref{pvar} that 
\begin{equation}
{d^{2}\phi\over{dy^{2}}} = - {1\over 2} \left({d\phi\over 
dy}\right)^{2}.
\end{equation}
It is now necessary to be careful about the limits of integration.  
Similar care was needed in the analysis of Ref.~\cite{bialek}, 
outside the region where the data $\{x_{i}\}$ lies.  It is 
interesting to note that this care in the boundary terms is 
unnecessary at $N=\infty,$ in accord with the fact that any finite 
data set will clearly not indicate the true limits of the probability 
distribution.
Suppose that $x$ ranges from $x_{-}$ to $x_{+},$ then 
integrating  eq.~\ref{pvar} I find 
\begin{equation}
N - i\lambda - \int_{x_{-}}^{x_{+}}
dx ({1\over \sqrt{h}}\phi')' =0 \ ,
\end{equation} 
so since
\begin{equation}
\int_{y_{-}\equiv y(x_{-})}^{y_{+}\equiv y(x_{+})} dz \exp(-\phi(z)) = 1\ ,
\end{equation}
it follows that 
\begin{equation}
\exp(+\phi) = (y_{+}-y_{-}) {(y-c)^{2}\over(y_{-}-c)(y_{+}-c)}
\end{equation}
where $c$ is an arbitrary constant of integration.  $c$ is restricted 
only by $c>y_{+}>y_{-},$ or $y_{+}>y_{-}>c,$  since $\exp(-\phi)$ must
be positive.  This `cyclic' constraint on $y_{+},y_{-},c$ is a hint of
projective invariance, expected in this theory because the algebra of
infinitesimal reparametrizations of the line has a subalgebra 
isomorphic to $sl(2,{\bf R}),$ familiar territory for string theorists.

Finally, we need to determine $y(x),$ which satisfies
\begin{equation}
{(y_{+}-c)(y-y_{-})\over(y_{+}-y_{-})(y-c)} = \int_{x_{-}}^{x} 
\left[{1\over N} \sum_{i}\delta(t-x_{i})\right] dx.
\end{equation}
Observe that the cross ratio on the left is projectively invariant, 
{\it i.e.}, invariant under transformations of the form 
\begin{equation}
z \mapsto {az+b\over cz+d},
\end{equation}
with $a,b,c,d$ real.  This amounts to a three-parameter family of
equivalent solutions $y(x)$ determined by the data.  
This projective invariance can be fixed by setting $c=\infty,y_{+}=1$
and $y_{-}=0.$  In this case, $\phi=0!$  

This is quite remarkable.  What we have just found is that the 
saddlepoint probability distribution deduced from the data $\{x_{i}\}$
is given by a function $Q(x) = y'(x),$ where $y$ is that 
reparametrization of the interval $(x_{-},x_{+})$ such that 
$\{y(x_{i})\}$ are uniformly distributed between $0$ and $1,$ in other
words, $\phi(y)=0.$
This solution is manifestly reparametrization invariant, since the
solution has picked a {\it canonical} set of coordinates.  Given {\it 
any} reparametrization of the data, the {\it same} canonical 
coordinates will be found, hence the probability distributions 
deduced in the two sets of coordinates will automatically be 
reparametrization-covariant, as desired.  

I omit the remaining steps in the semi-classical analysis since they
are clearly explained in Ref.~\cite{bialek}.  The slight complication 
here is in explicitly treating the division by the volume of the 
diffeomorphism group, but this is easily handled in one dimension by
the Faddeev-Popov method.  

Of considerable interest is the generalization of the 
reparametrization invariant theory given above to the case of dimensions
much larger than one\cite{callan}.  
It seems  clear that we should not be integrating over
all possible metrics since there are entirely too many degrees of 
freedom in such an integration in contrast to integrating over
metrics in one, two, or three dimensions.  
A suitable constraint might be to integrate only over 
metrics with vanishing Weyl curvature, which are metrics
such that there exist coordinates in which the metric is proportional 
to a constant matrix.  In fact, counting local 
degrees of freedom, it would appear that it is unnecessary to 
even introduce a function $\phi$ into the analysis for dimensions
larger than three;  there are exactly 
enough degrees of freedom in metrics with vanishing Weyl curvature to 
leave a local scalar degree of freedom after dividing by the diffeomorphism 
group.  This is the same count that was used in one dimension, so it
suggests that the correct generalization to higher dimensions is 
given by 
\begin{equation}
P[h] \equiv {1\over Z} \exp\left(-\int d^{D}x \sqrt{h(x)} 
R(h_{\alpha\beta}(x))\right) \delta\left(1-\int d^{D}x 
\sqrt{h(x)}\right),
\end{equation}
where $R(h_{\alpha\beta})$ is the Ricci scalar constructed out of the
metric $h_{\alpha\beta},$ constrained to be a metric of vanishing 
Weyl curvature, and $\sqrt{h}\equiv det^{1/2}(h_{\alpha\beta})$ as 
usual.  However, the analysis of the equations that follow
from this ansatz is more complicated because the constraint
of vanishing Weyl curvature contributes a source term in the equation
analogous to eq.~\ref{hvar} above.  Further,  
while the weak-field limit of $R$ seems to give the correct damping
of gradients, it is unclear if there is a positivity theorem that
could be proved for $R(h_{\alpha\beta}),$ 
subject to the constraint of vanishing Weyl 
curvature.  This is under investigation.

I am grateful to Curt Callan for helpful conversations. 
This work was supported in part by NSF grant PHY96-00258.

\end{document}